\documentclass[11pt]{article}

\usepackage[a4paper,margin=1in]{geometry}
\usepackage{amsmath,amssymb,amsfonts}
\usepackage{amsthm}
\usepackage{mathrsfs}
\usepackage{bm}
\usepackage{hyperref}
\usepackage{enumerate}
\usepackage{cite}

\usepackage{graphicx}
\usepackage{caption}
\usepackage{subcaption}

\theoremstyle{plain}

\theoremstyle{definition}

\theoremstyle{remark}

\begin{document}
	
	\title{\textbf{A Global Optimal Theory of Portfolio beyond R-$\sigma$ Model} }%\\
%		\large Subtitle if Necessary}
	\author{
	Yifan Liu\thanks{liuyfsysu@163.com}   \   and\   %\\
	Shi-Dong Liang\thanks{stslsd@mail.sysu.edu.cn}\\	
		\small School of Physics, Sun Yat-Sen University\\
		\small Guangzhou, China
	}
	
	\date{\today}

	\maketitle
	
	% =========================
	% 摘要
	% =========================
	\begin{abstract}
The deviation of the efficient market hypothesis (EMH) for the practical economic system
allows us gain the arbitrary or risk premium in finance markets.
We propose the triplet $(R,H,\sigma)$ theory to give the local and global optimal
portfolio, which generalize from the $(R,\sigma)$ model.
We present the formulation of the  triplet $(R,H,\sigma)$ model and give the Pareto optimal solution as well as comparing it with the numerical investigations for the Chinese stock market. We define the local optimal weights of the triplet $(\mathbf{w}_{R},\mathbf{w}_{H},\mathbf{w}_{\sigma})$, which constructs
the triangle of the quasi-optimal investing subspace such that we further define the
centroid of the triangle or the incenter of the triangle
as the optimal investing weights, which optimizes the mean return, the arbitrary or risk
premium and the volatility risk. By investigating numerically the Chinese stock market as an example we demonstrate the validity of the formulation and obtain the global optimal
strategy and quasi-optimal investing subspace. The theory provides an efficient way to design the portfolio for different style investors, conservative or aggressive investors, in finance market to maximize the mean return and arbitrary or risk premium with a small volatility risk.
	\end{abstract}
	
	\tableofcontents
	
	% =========================
	\section{Introduction}
	% =========================
	
As one of fundamental classic finance theories, Efficient Market Hypothesis (EMH) has been questioned by some practical economic systems, such as the stock markets.\cite{De,Loughran,Chen} Some novel financial phenomena, such as momentum and contrarian effect, mean reversion effect indicate stock returns follows the stable Levy distribution rather than standard Gaussian distribution. \cite{ANTEGNA} The stock price movements are influenced by the investors' psychology. Edgar E. Peters proposed Fractal Market Hypothesis (FMH) to analyze the capital markets by non-linear and fractal theories and methods. \cite{Peters} The key features of FMH are that the time sequences of the asset prices show the self-similarity and has the long-term correlation. \cite{Jegadeesh,Gebhardt,Drew} In principle, the Hurst exponent $(0<Hurst<1)$ can be used as an index to measure quantitatively the long-term correlation. Some studies showed that Hurst exponent could be regarded as a kind of risk premium: the bigger is, the more risk or speculate premium one asset contains, and investors could get the risk premium by using trend or reversal strategies.\cite{Wei,Gao}

Using the fractal and multifractal theories to analyze stock market become a branch of econophysics. Markwitz proposed a mean-variance $(R,\sigma)$ model based on the standard Gaussian process for the optimization asset allocation, which guides the investor to maximize the mean return and minimize the risk. \cite{Markowitz} However, some practical stock markets deviate the standard Gaussian process and exhibit fractal characteristic, \cite{Peters,Peters2} which provides some chance to optimize their asset allocation by using these fractal behaviors. The question is how to design an optimal investing strategy or how to select the asset allocation or portfolio based on the fractal stochastic time sequences and their historical data.

In this paper, we generalize the mean-variance model of portfolio to a triplet mathematical model $(R,\sigma,H)$ of portfolio, which contains the mean return, risk and speculate premium. Based on this model we give the Pareto optimal solution of portfolio and the global optimal solution of portfolio as well as the quasi-optimal subspace for different investment options. We also propose two practical strategies to find out the optimal or relative optimal investing schemes of portfolio for different investors. In Sec. II, we will present the triplet mathematical model of portfolio $(R,\sigma,H)$ and its Pareto optimal solution. We will define the global optimal solution and the quasi-optimal subspace of portfolio in Sec. III. We propose two practical strategies to find out the global and quasi-optimal solutions of portfolio in Sec. V. Finally, we will give the conclusion.

	% =========================
	\section{$(R,\sigma,H)$ model of portfolio and its Pareto solution}
	% =========================

The main goal of portfolio is to maximize the return and minimize the risk.
Based on the EMH, the higher return, the higher risk, which implies
that there does not exist the arbitrage space. However,
the practical economic systems deviate the EMH, namely the stochastic processes
deviate the standard Gaussian process, which induces some arbitrage space.
Since the Hurst exponent measures the deviation of the stochastic process from
the standard Gaussian process, we take the Hurst exponent into account
to set up a triplet mathematical model $(R, \sigma, H)$, where $R$ denotes the expected
returns and its corresponding variance $\sigma$. $H$ is the Hurst exponent, which
provides some information for the risk or speculate premium.
This model generalizes the mean-variance model $(R, \sigma)$ by Markwitz.\cite{Markowitz}

Let us set $\{r_{i}(\tau_{k})\}$ be the time sequence of the expected returns for the asset $i$ at the time interval $\tau_k=t_k-t_{k-1}$, where $i=1,2,...N$ labels different asset and $k=1,2,...M$ labels the time interval.
The time-mean return of the asset vector in the whole period is expressed as
\begin{equation}\label{TR}
	\mathbf{R}:=\left(R_{1},R_{2},\cdots,R_{N}\right) =\frac{1}{M}\sum_{k=1}^{M}\mathbf{r}(\tau_{k})
\end{equation}
where
$R_{i}=\left(\sum_{k=1}^{M}r_{1}(\tau_{k}),\sum_{k=1}^{M}r_{2}(\tau_{k}),
\cdots,\sum_{k=1}^{M}r_{M}(\tau_{k})\right)/M$.

Suppose that the portfolio weights for different assets are denoted by a vector,
$\mathbf{w}=\{w_i\}\in \Omega$ with $\sum_{i=1}^{N}w_i=1$, and $w_i\geq 0$, where $\Omega$ is the sample space in the probability space of the portfolio.
The total return of the portfolio can be expressed as
\begin{equation}\label{rj}
	\left\langle R\right\rangle=\mathbf{w}^{\top}\mathbf{R}
\end{equation}
with its corresponding variance
\begin{equation}\label{V1}
	\sigma^2(\mathbf{w})=\left\langle\left(R_{i}-\left\langle R\right\rangle\right)^{2}\right\rangle
	=\mathbf{w}^{\top} \mathbf{C}\mathbf{w}
	%=\frac{1}{M}\sum_{j=1}^{M}\sum_{i=1}^{N}w_{i}\left(r_{i}(\tau_{j})-R_{M}\right)^{2}
\end{equation}
where
\begin{equation}\label{C}
	\mathbf{C}=(\sigma_{ij})
\end{equation}
with
\begin{equation}\label{S}
	\sigma_{ij}=\langle R_{i}R_{j}\rangle
	-\langle R_{i}\rangle\langle R_{j}\rangle
\end{equation}
is the correlation matrix of the return.

For a given asset $i$, its return is a stochastic series
with the time step $k$, $\{r_{i}(\tau_{k})\}$. Suppose that the time series $\{r_{i}(\tau_{k})\}$
deviate the standard Gaussian process, which can be measured by the Hurst exponent.
When $H_i=1/2$ means the time series $\{r_{i}(\tau_{k})\}$ runs with the standard Gaussian process, and $H_i>1/2$ predicts a persistency correlation, and
$H_i<1/2$ predicts an anti-persistency correlation.
The deviation of the standard Gaussian process
provides some arbitrage space for investors.
The investigation of the Chinese stock market indicates that the stock series
deviates the standard Gaussian process and its Hurst exponent is around 0.6\cite{Shi,Han}.
This implies that the stock market runs a fractal Browan process and
there exists some arbitrage space or risk premium in the stochastic
market. We may assume that the investors favor the non-standard Gaussian process
in the stochastic market because the non-standard Gaussian process implies the existence of
some information for speculation.
The Hurst exponent can be calculated by (see Appendix)
\begin{equation}\label{HEp}
	H_i=\frac{\log(F_{i}(s)/C)}{\log s}
\end{equation}
where $F_{i}(s)$ is the fluctuation function of the stochastic sequences.
We introduce the mean Hurst exponent,
\begin{equation}\label{MH}
	\langle H\rangle=\sum_{i=1}^{N}w_{i}H_i=\mathbf{w}^{\top}\mathbf{H}
\end{equation}

In general, it can be assumed that any asset time series is a non-Markov chain and
non-standard Gaussian process. The non-Markov properties of the stochastic
sequences promise the predicability of the stochastic sequences based on the prior
information of the sequences. The non-standard Gaussian properties of the stochastic
sequences provide an arbitrage space for investors.

The question is how investors design an optimal asset allocation or portfolio based on the prior information of the stochastic sequences.
In general we can define the optimal weight of the portfolio.

\textbf{Definition}: The Pareto optimal weight of the portfolio is defined by
\begin{equation}\label{OW}
	\left\{\mathbf{w}^{*}\in\Omega^{N} \left| \max_{\mathbf{w}^{*}\in\Omega}R(\mathbf{w}), \right.
	\min_{\mathbf{w}^{*}\in\Omega}\sigma^{2}(\mathbf{w}),
	\left(\mathbf{w}^{*\top}\mathbf{H}\right)\geq \frac{1}{2} ,
	\mathbf{1}^{\top}\mathbf{w}^{*}=1 \right\}
\end{equation}
where $\mathbf{1}^{\top}=(1,1,\cdots,1_N)$.

The main goal of investors is to design a weight for portfolio to maximize the return $\max_{\mathbf{w}^{*}\in\Omega}R_{M}$ and to minimize the uncertainty and risk.
The constraint $\left(\mathbf{w}^{*\top}\mathbf{H}\right)\geq \frac{1}{2}$ means that we can search for the bigger arbitrage space for investors.
The optimal asset portfolio problem is a multi-objected optimization problem from the mathematical point of views. Thus, the portfolio space
can be generalized to three-dimensional space Return-Risk-Arbitrage (RRA) from
the conventional two-dimensional space Return-Risk (RR).\cite{Clark}
We may define the Lagrangian function by\cite{Clark}
\begin{equation}\label{Lf}
	L= R(\mathbf{w})-\sigma^2(\mathbf{w})
	+\mathbf{\lambda}(\mathbf{g}(\mathbf{w})-\mathbf{b})
\end{equation}
where
$\mathbf{\lambda}=(\lambda_{1},\lambda_{2})$ is the Lagrangian multiplier vector and
\begin{eqnarray}
	R(\mathbf{w}) &=& \mathbf{w}^{\top}\mathbf{R} \\
	\sigma^2(\mathbf{w}) &=& \mathbf{w}^{\top} \mathbf{C}\mathbf{w} \\
	\mathbf{g}(\mathbf{w}) &=& \left(\left(\mathbf{w}^{\top}\mathbf{H}\right), \mathbf{w}\right)^{\top} \\
	\mathbf{b} &=& \left(\frac{1}{2}, 1\right)^{\top}
\end{eqnarray}

The Lagrangian function as the objective function can be maximized by the Kuhn-Tucker theorem.\cite{Clark} The Kuhn-Tucker conditions are
\begin{eqnarray}
	&& \mathbf{R}-2\mathbf{C}\mathbf{w}
	+\lambda_{1}\mathbf{H}+ \lambda_{2}\mathbf{1}\leq\mathbf{0}\\
	&& \mathbf{w}^{\top}\mathbf{H}-\frac{1}{2}\geq 0 \\
	&& \mathbf{w}^{\top}\mathbf{1}-1= 0 \\
\end{eqnarray}
By solving the Kuhn-Tucker conditions, we obtain
\begin{equation}\label{OW}
	\mathbf{w}^{*}=\frac{1}{2}\mathbf{C}^{-1}\left(\mathbf{R}
	+\lambda_{1}\mathbf{H}+\lambda_{2}\mathbf{1}\right )
\end{equation}
where
\begin{equation}\label{LD}
	\left(
	\begin{array}{c}
		\lambda_{1} \\
		\lambda_{1}
	\end{array}
	\right)
	=\frac{1}{\det \mathbf{Q}}\left(
	\begin{array}{c}
		\alpha \mathbf{1}^{\top}\mathbf{C}^{-1}\mathbf{1}
		-\beta\mathbf{H}^{\top}\mathbf{C}^{-1}\mathbf{1}\\
		-\alpha\mathbf{1}^{\top}\mathbf{C}^{-1}\mathbf{H}
		+\beta\mathbf{H}^{\top}\mathbf{C}^{-1}\mathbf{H}
	\end{array}
	\right)
\end{equation}
with
\begin{equation}\label{Q}
	\mathbf{Q}
	=\left(
	\begin{array}{cc}
		\mathbf{H}^{\top}\mathbf{C}^{-1}\mathbf{H} & \quad \mathbf{H}^{\top}\mathbf{C}^{-1}
		\mathbf{1} \\
		\mathbf{1}^{\top}\mathbf{C}^{-1}\mathbf{H} & \mathbf{1}^{\top}\mathbf{C}^{-1}\mathbf{1}
	\end{array}
	\right)
\end{equation}
\begin{eqnarray}\label{ab}
	\alpha &=& 1- \mathbf{H}^{\top}\mathbf{C}^{-1}\mathbf{R}
	\\
	\beta &=& 2- \mathbf{1}^{\top}\mathbf{C}^{-1}\mathbf{R}
\end{eqnarray}

The formulation on portfolio we obtain from ($\ref{OW}$) to ($\ref{ab}$) provides a
efficient way to search for the Pareto optimal weight of portfolio for investors.

	% =========================
	\section{Global optimal solution and Quasi-optimal subspace}
	% =========================
However, the Pareto optimal solution is based on the constraint
$\left(\mathbf{w}^{*\top}\mathbf{H}\right)\geq \frac{1}{2}$. There should be either
some bigger arbitrage space for the aggressive investors or concrete space for
the conservative investors. Thus, we define the local optimal weight of portfolio.

\textbf{Definition of Local optimum}:
The weights $\mathbf{w}_{R},\mathbf{w}_{\sigma},\mathbf{w}_{H}\in\Omega$
are local optimal respectively if they satisfy
\begin{equation}\label{WLO}
	(\mathbf{w}_{R},\mathbf{w}_{\sigma},\mathbf{w}_{H}):=
	\left(\max_{\mathbf{w}\in\Omega}\{\langle R\rangle\},\min_{\mathbf{w}\in\Omega}\{\sigma\},
	\max_{\mathbf{w}\in\Omega}\{H\}\right)
\end{equation}
where
$\mathbf{w}_{R}:=\{w^{R}_{i}\}_{\max_{\mathbf{w}\in\Omega}\{\langle R\rangle\}}$, $\mathbf{w}_{\sigma}:=\{w_{i}^{\sigma}\}_{\min_{\mathbf{w}\in\Omega}\{\sigma\}}$, and $\mathbf{w}_{H}:=\{w_{i}^{H}\}_{\max_{\mathbf{w}\in\Omega}\{H\}}$ .

In general $\mathbf{w}_{R}\neq\mathbf{w}_{\sigma}\neq\mathbf{w}_{H}$ and they
could be competed each other for different benefits, which give the extreme values for
investors to estimate their expected return, the biggest arbitrage space and
the worst risk. For convenience we define a quasi-subspace within the whole
investing space as a domain of investment for different strategies of different style investors. Suppose that $(\mathbf{w}_{R},\mathbf{w}_{\sigma},\mathbf{w}_{H})$
forms a triangle in the sample space without losing generality.

\textbf{Definition of Quasi-optimal subspace $\Omega_T$}: The subset $\Omega_T\subset\Omega$ is the quasi-optimal subspace if it is within
the triangle consisted of the vertex  $\left(\mathbf{w}_{R},\mathbf{w}_{\sigma},\mathbf{w}_{H}\right)$ in the sample space.

This quasi-optimal subspace provides a practical domain of portfolio in the investing space
for different investing style investors. The aggressive investors may
select the weights close to $\mathbf{w}_{R}$ and the conservative investors
may select the weights close to $\mathbf{w}_{\sigma}$.

What is the global optimal weight for investors?
If we assume to share equally the three factors, return, risk and arbitrage,
the global optimal weight should close the three local
optimal weights as much as possible. Therefore,
We define the global optimal weight $\mathbf{w}_{O}$ by

\textbf{Definition of Global optimum $\mathbf{w}_{O}$}: The weight of the portfolio $\mathbf{w}_{O}=\{w_i\}_O\in\Omega$ is global optimal if it satisfies
$\left( \min_{\mathbf{w}\in\Omega} d_{O,R},
\min_{\mathbf{w}\in\Omega} d_{O,\sigma},
\min_{\mathbf{w}\in\Omega} d_{O,H}\right)$ simultaneously.
where $ d_{O,\alpha}$ is the distance between
$\mathbf{w}_{O}$ and $\mathbf{w}_{\alpha}$, where $\alpha=R,\sigma, H$.

This global optimal weight can be regarded as a Pareto solution of the triplet game.
The weight can be viewed as a vector. Thus, we can define the distance between two local weights by the norm of the difference between two weight vectors in Euclidian space.

\textbf{Definition of Distance}:
The distance between two vectors $\mathbf{w}_{\alpha},\mathbf{w}_{\beta}$ is defined by
the norm in Euclidian space,
\begin{equation}\label{dd}
	d_{\alpha,\beta} = \|\mathbf{w}_{\alpha}-\mathbf{w}_{\beta}\|
\end{equation}
Based on the definition of the distance between two local weights, we can define the  distance for any investing weight away from the global optimal weight.
\begin{equation}
	d_{\mathbf{w},O}:=
	\| \mathbf{w}-\mathbf{w}_{R}\|+
	\| \mathbf{w}-\mathbf{w}_{\sigma}\| + \| \mathbf{w}-\mathbf{w}_{H}\|
\end{equation}
Thus, the basic idea to find out the global optimal weight is to minimize $d_{O}:=\min_{\mathbf{w_O}\in\Omega} d_{\mathbf{w},O}$.
It implies the global optimal solution is within the triangle constructed by the
local optimal weights, $\mathbf{w}_{O}\in \Omega_T$. Thus, we propose two schemes to find out the global optimal solution of portfolio.

\textbf{Global optimal solution}: The global optimal solution of portfolio
is  either

(1) the centroid of the triangle $\Omega_T$,
\begin{equation}\label{WO1}
	w_{i,O}^{(c)}=\frac{1}{3}(w_{i}^{R}+w_{i}^{\sigma}+w_{i}^{H})
\end{equation}

or

(2) the incenter of the triangle $\Omega_{IC}$,
\begin{equation}\label{WO2}
	w_{i,O}^{(ic)}=\frac{d_{\sigma,H}w_{i}^{R}+d_{R,H}w_{i}^{\sigma}
		+d_{R,\sigma}w_{i}^{H}}{d_{\sigma,H}+d_{R,H}+d_{R,\sigma}}
\end{equation}
where $d_{\alpha,\beta}$ is the distance between the weights of $\alpha$ and $\beta$.

\textbf{Quasi-optimal subspace}: The quasi-optimal subspace is either

(1) the triangle $\Omega_T$ constructed by the local optimal weights, $\left( \mathbf{w}_{R}, \mathbf{w}_{H},\mathbf{w}_{\sigma}\right)$.

or

(2) the incenter of the triangle $\Omega_{IC}$, which radius is
\begin{equation}\label{WOI}
	r_{I}=\sqrt{\frac{(s-d_{\sigma,H})(s-d_{R,H})(s-d_{R,\sigma})}{s}}
\end{equation}
where
\begin{equation}\label{AL}
	s=\frac{d_{\sigma,H}+d_{R,H}+d_{R,\sigma}}{3}
\end{equation}

The global optimal solution and the quasi-optimal subspace provide a lot of investing strategies or schemes for different style investors to gain their expected mean returns.
The optimal solutions of portfolio, the centroid of the triangle and the incenter of the triangle, balance the mean return, arbitrary premium and volatility risk, which can be regarded a Pareto solution of the triplet game.
The two quasi-optimal subspaces, the triangle and the incircle of the triangle,
can be regarded as two efficient subspace in the investing space, which provide
a lot of choices for differential style investors to play their investing strategies.
The aggressive investor may choose the weights near the local optimum of the mean return $\mathbf{w}_{R}$ or near both of $\mathbf{w}_{R},\mathbf{w}_{H}$ inside of the
triangle. The conservative investors may choose the weights near the local optimum of the volatility $\mathbf{w}_{\sigma}$ inside of the triangle.

It should be remarked that the quasi-optimal subspace $\Omega_T$ could depend on the
time interval $\tau$ we choose. We denote the quasi-optimal subspace $\Omega_{T}^{\tau}$
for given time interval $\tau$. In order to reduce the uncertainty induced by
the time interval we choose, we propose two schemes to enhance the predicability
of the optimal strategies. One is that we make the average of the local optimal
weights for different time interval $\tau$, which is defined by
\begin{eqnarray}\label{AWWW}
	\mathbf{w}_{R}^{\tau}:&=& \frac{1}{N_{\tau}}\sum_{n}^{N_{\tau}}\mathbf{w}_{R}^{\tau_n} \\
	\mathbf{w}_{\sigma}^{\tau}:&=& \frac{1}{N_{\tau}}\sum_{n}^{N_{\tau}}\mathbf{w}_{\sigma}^{\tau_n} \\ \mathbf{w}_{H}^{\tau}:&=& \frac{1}{N_{\tau}}\sum_{n}^{N_{\tau}}\mathbf{w}_{H}^{\tau_n}
\end{eqnarray}
where $N_{\tau}$ is the number of the time intervals we choose.
The averages of the local optimal weights for different time intervals
as an effective local optimal weights can reduce the uncertainty of the time intervals
we choose. Based on this effective local optimal weights we can
define the effective quasi-optimal subspace by the triangle consisted of the vertexes
$(\mathbf{w}_{R}^{\tau},\mathbf{w}_{\sigma}^{\tau},\mathbf{w}_{H}^{\tau})$.

The other way to reduce the uncertainty from the time interval is that we
define the effective quasi-optimal subspace $\Omega_T$ by the overlap of the quasi-optimal subspaces for different given time interval $\tau$.
\begin{equation}\label{O}
	\Omega_T^{eff}=\bigcap_{\tau}\Omega_T^{\tau}
\end{equation}

The averages of the local optimal weights enhance the predicability of the optimal
strategies. The effective quasi-optimal subspace provides a way to shrink the
effective investing space of portfolio.

\begin{figure}
	\centering
	\includegraphics[width=0.7\textwidth]{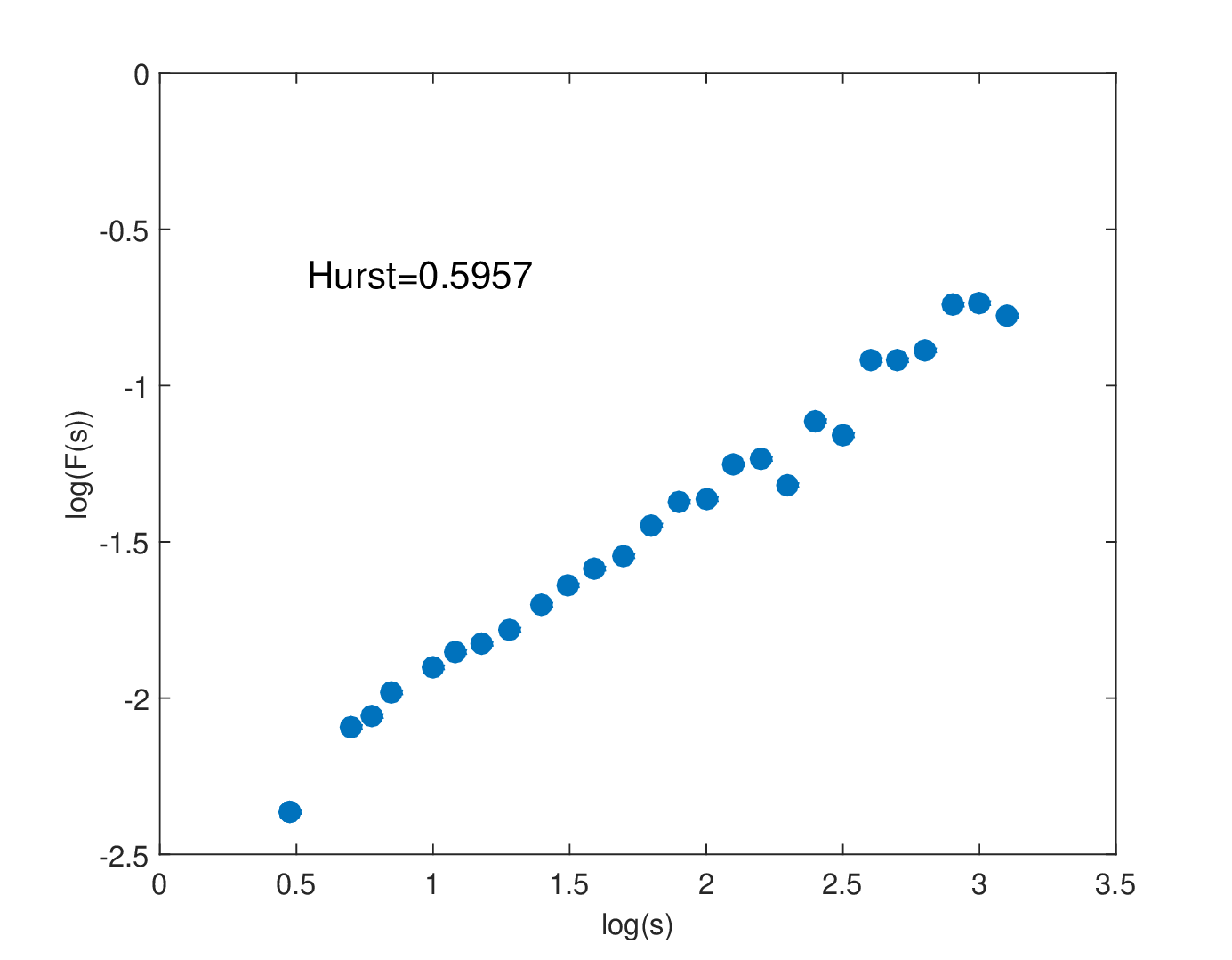}
	\caption{(Color online)The $\log F(s)-\log(s)$ of the index subseries of the Chinese stock market.}
	\label{fig1}
\end{figure}

	% =========================
	\section{Applications in Stock market}
	% =========================
\subsection{Fractal characteristic in Stock market}
The stock market is a stochastic time sequences.
We analyze the basic characteristic of the Chinese stock market based on the Hurst exponent method. We use $1292$ daily data of the Shanghai (securities) composite index from Jan. 4 2013 to Apr. 27 2018. The subsequences length $s$ of the stock index log-return series is from $s=5$ to $1292$. By plotting $\log(F(s))-\log(s)$ (see Fig. $\ref{fig1}$) with the linear fitting. We obtain the Hurst exponent of the Shanghai (securities) composite index $0.5957$, which means that the Chinese stock market does not follow the geometric Brownian motion, but the fractal Brownian motion.

\begin{figure}
	\centering
	\includegraphics[width=0.7\textwidth]{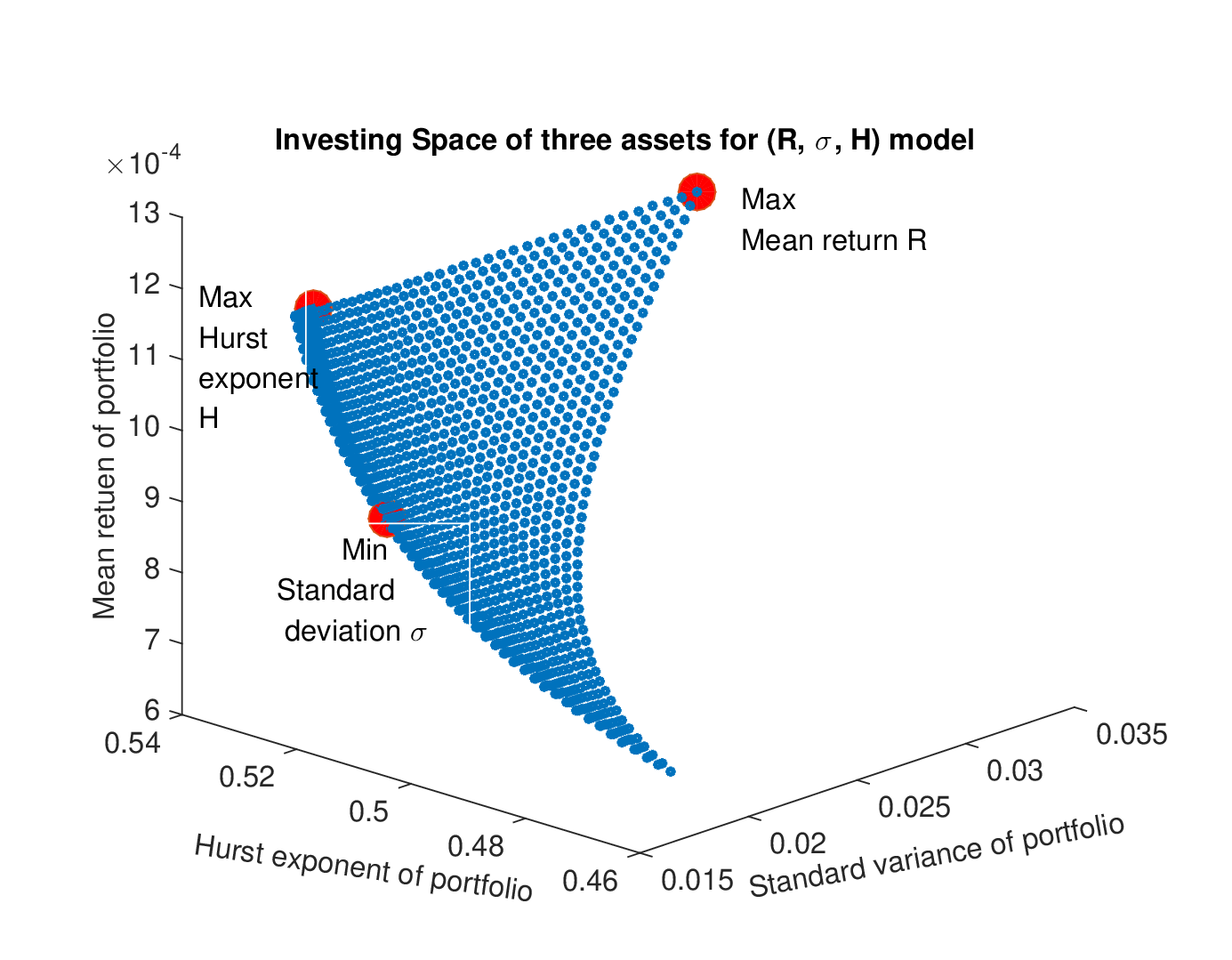}
	\caption{(Color online):The investing space based on the $(R,\sigma,H)$ model, where
		the time interval is one day.
		The red points are the local optimal weights for $(\max R, \min\sigma, \max H)$.}
	\label{fig2}
\end{figure}

The fractal characteristic emerging in Chinese stock market
implies that there exists the speculate or risk premium for investors.
The rational investors can optimize their asset allocation based on maximizing $|H-0.5|$ .
This provides some chance to optimize the portfolio based on the above formulation of  portfolio.

\begin{table}
	\caption{Optimal Weights and Their Mean Returns}
	\begin{tabular}{|c |c |c | c|}
		\hline \hline
		LOW$^{1}$ & $\mathbf{w}_{R}(0,0,1)$ & $\mathbf{w}_{H}(0,1,0)$ & $\mathbf{w}_{\sigma}(0.40,0.48,0.12)$ \\
		DMR$^{2}$ & 0.12\% & 0.11 \% & 0.094\% \\
		VR$^{3}$ &0.0328 & 0.0196 & 0.0160  \\\hline
		OW$^{4}$ &  $\mathbf{w}^{c}_{O}(0.49,0.37,0.14)$ & $\mathbf{w}^{ic}_{O}(0.56,0.26,0.18)$
		& $\mathbf{w}^{*}_{O}(0.5071,0.3003,0.1925)^{6}$ \\
		ODMR$^{5}$  & 0.090\% & 0.087\% & 0.089\% \\
		VR$^{3}$ & 0.0161 & 0.0178& 0.0165   \\
		\hline \hline
	\end{tabular}
	$^{1}${LOW: Local Optimal Weight}; $^{2}${DMR: Daily Mean Return};
	$^{3}${VR: Volatility Risk}; \\
	$^{4}${OW: Optimal Weight:$\mathbf{w}^{c}$: Centroid of triangle;
		$\mathbf{w}^{ic}$: Incenter of triangle}\\
	$^{5}${ODMR: Optimal Daily Mean Return};
	$^{6}${Calculated by Eqs. ($\ref{OW}$)-($\ref{AWWW}$)}
\end{table}
	
\subsection{Optimal strategic space}

The $(R,\sigma,H)$ model we generalized provides more opportunities to gain the
arbitrage or risk premium for investors. Let us analyze some stock running
as an example to demonstrate how our method works.
We investigate the strategic space of three stocks,
Yunnan Baiyao (000538.SZ), Guizhou Maotai (600519.SH) and Iflytek (002230.SZ),
in Chinese stock market based on the formulation in Eqs. ($\ref{OW}$)-($\ref{AWWW}$) and the numerical method in Eq. ($\ref{WLO}$). The data we investigated are from Jan. 4, 2013 to April 27, 2018 and the time intervals we chose are one day.

\begin{figure}
	\centering
	\includegraphics[width=0.7\textwidth]{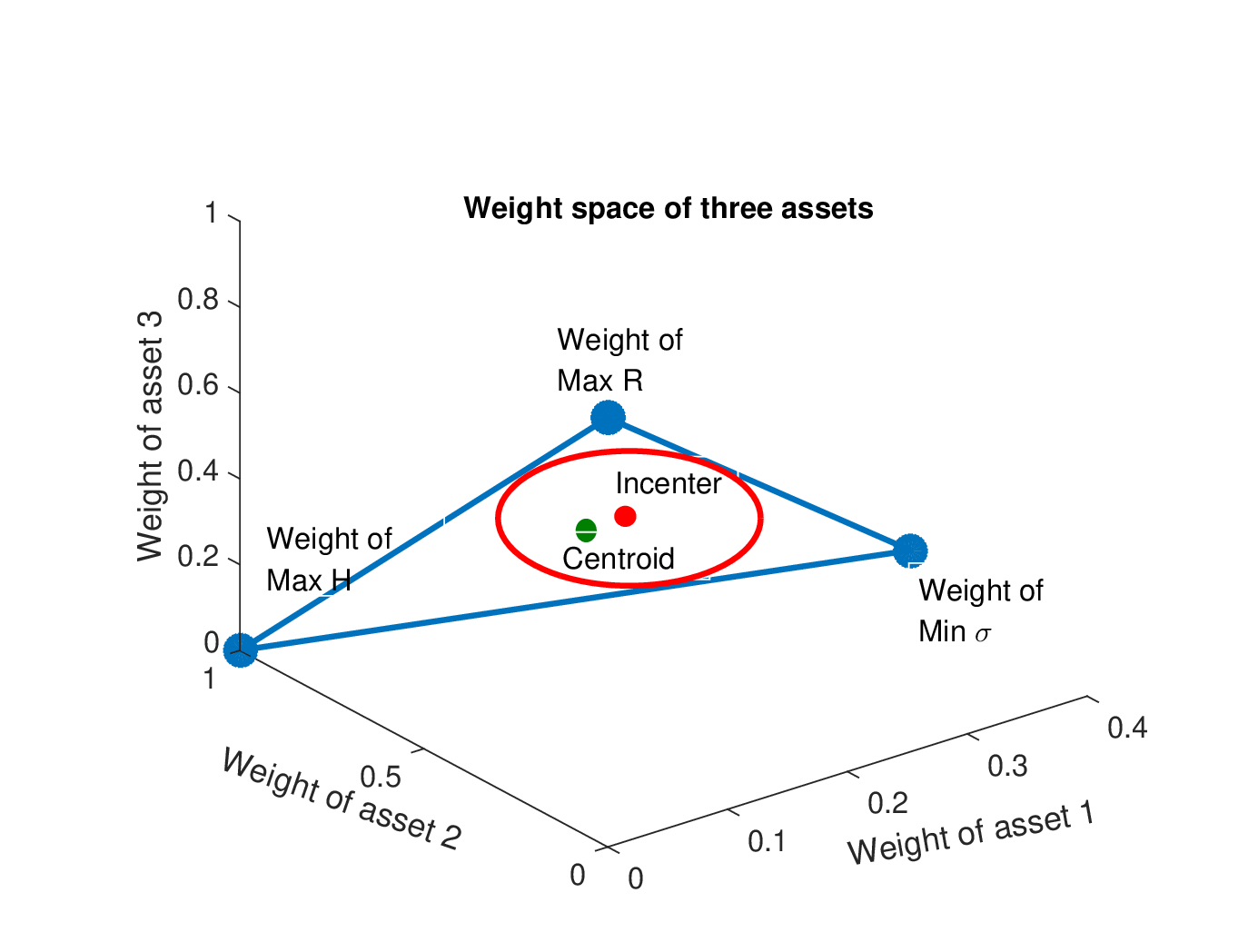}
	\caption{(Color online):  The quasi-optimal investing subspace of $(R,\sigma,H)$ model.}
	\label{fig3}
\end{figure}

The investing space of the $(R,\sigma,H)$ model for these three stocks in Chinese stock market is shown in Fig. ($\ref{fig2}$), in which the red points are the local optimal weights for $(\max R, \min\sigma, \max H)$. These points provide a guideline for different style investors, aggressive or conservative investors.

The local optimal weights, optimal weights and their daily mean returns are
calculated by the formulation in Eqs. ($\ref{OW}$)-($\ref{AWWW}$) and by the numerical results directly from the definition ($\ref{WLO}$), which are listed in the table I.

\begin{table}
	\caption{The Averages of Optimal Weights and Their Mean Returns}
	\begin{tabular}{|c |c |c | c|}
		\hline \hline
		ALOW$^{1}$ & $\mathbf{w}_{R}(0,0,1)$ & $\mathbf{w}_{H}(0,0.9,0.1)$ & $\mathbf{w}_{\sigma}(0.51,0.42,0.07)$ \\
		ADMR$^{2}$ & 0.12\% & 0.11 \% & 0.088\% \\
		SE$^{3}$ & 0.0328 & 0.0185 & 0.0161  \\\hline
		AOW$^{4}$ &  $\mathbf{w}^{c}_{O}(0.44,0.39,0.21)$ & $\mathbf{w}^{ic}_{O}(0.50,0.29,0.17)$
		& $\mathbf{w}^{*}_{O}(0.5373,0.2585,0.2042)^{6}$ \\
		AODMR$^{5}$  & 0.093\% & 0.090\% & 0.088\% \\
		SE$^{3}$ & 0.0161 & 0.0166 & 0.0167  \\\hline
		\hline
	\end{tabular}
	$^{1}${ALOW: Average of Local Optimal Weight of the time interval from 1 to 10 days}\\
	$^{2}${ADMR: Average of Daily Mean Return};
	$^{3}${SE: Standard Errors for different time intervals}; \\
	$^{4}${AOW: Averages of Optimal Weight:
		$\mathbf{w}^{c}$: Centroid of triangle; $\mathbf{w}^{ic}$: Incenter of triangle}\\
	$^{5}${AODMR: Average of Optimal Daily Mean Return};
	$^{6}${Calculated by Eqs. ($\ref{OW}$)-($\ref{AWWW}$)}\\
	
\end{table}

It can be seen that the daily returns of the local optimal weights is best
at the local maximum mean return point $\mathbf{w}_{R}$, the second at the local maximum
Hurst exponent point $\mathbf{w}_{H}$ and the lowest at the minimum variance point$\mathbf{w}_{\sigma}$.
However, the volatility risk is inverse the biggest for $\mathbf{w}_{R}$, the lowest
for $\mathbf{w}_{\sigma}$. This make sense. The daily mean return of optimal weight at the centroid of the triangle is better that at the incenter of the triangle, which
implies that the optimal weight estimated by the centroid of triangle is better than
the weight by the incenter of triangle from the daily mean return point of views.
Both of them are close to the analytic result by the formulation ($\ref{OW}$)-($\ref{AWWW}$).

In order to examine the effect of the time interval we investigate the
local optimal weights, optimal weights and their mean returns by the time intervals from one to ten days. The average results in this period are listed in the table II.

The average results show that the local optimal weights are very similar to the one-day results and their corresponding mean returns have the same order, namely
$R(\mathbf{w}_{R})>R(\mathbf{w}_{H})>R(\mathbf{w}_{\sigma})$ and
$R(\mathbf{w}^{c}_{O})>R(\mathbf{w}^{ic}_{O})$. The standard errors for different
time intervals from 1 to 10 days are around $0.016\sim 0.033$, which implies
that the time intervals in the short periods are not very sensitive for our
predictions.

The $(R,\sigma,H)$ model provides an efficient way to find out the local and global
optimal investing schemes for different style investors, conservative and aggressive
investors. In principle the triangle based on the local optimal weights
can be regarded a quasi-optimal subspace for investors, which provides a various
portfolio for investors. The centroid or incenter of the triangle can be regarded as
the optimal portfolio, which balance the mean return, the arbitrary or risk premium and
volatility risk.

\section{Conclusions}
There have been many evidences demonstrating the practical economic system, such as
the stock market, deviates the efficient market hypothesis (EMH), which induces
two fundamental issues. How do we avoid the risk or gain the arbitrary or risk premium
in finance markets? We propose the triplet $(R,\sigma,H)$ theory to give the local and global optimal portfolio, which generalize from the $(R,\sigma)$ model.\cite{Markowitz}
We give the formulation to give the Pareto optimal solution and compare
it with numerical investigations for the Chinese stock market to demonstrate the validity of
this formulation. Based on this theory we define the local optimal weights of the triplet $(\mathbf{w}_{R},\mathbf{w}_{H},\mathbf{w}_{\sigma})$, which consist of
the triangle of the quasi-optimal investing subspace such that we give the
global optimal solution of this triplet game, centroid of the triangle or the incenter of the triangle as the optimal investing weights, which optimizes the mean return, the arbitrary or risk premium with a small volatility risk.

The theory provides an efficient way to design the portfolio for different style investors, conservative or aggressive investors, in finance market to maximize the mean return with
a relative small volatility risk. This idea could be generalize to other economic systems to find out the Pareto solution.

%\section*{Declarations}

%Some journals require declarations to be submitted in a standardised format. Please check the Instructions for Authors of the journal to which you are submitting to see if you need to complete this section. If yes, your manuscript must contain the following sections under the heading `Declarations':

\section{Appendices}

\subsection{Hurst exponent}
Hurst exponent is introduced by Hurst to describe the fractal Brownian motion and stochastic time series as well as their long-term correlation. Hurst exponent is within three different regions:

(1) $0<H<0.5$ means that the stochastic time series is anti-persistency, namely the past and future series are negative correlation. The sequence contains a sudden reversal property.

(2) $H=0.5$ means that the time series follows independent stochastic distribution, namely the standard Brownian motion. The past and future series are independent.

(3) $H<0.5<1$ means that the time series is persistency, namely the past and future series are positive correlation.

\subsection{Calculating method of Hurst exponent}
In general the Hurst exponent of the time series can be calculated by the Rescale Range Analysis (R/S) or the Detrended Fluctuation Analysis (DFA). Here we use the DFA method to calculate Hurst exponent.

For a time series $x:=\{x(\tau_k)\}$, where $\tau_k=t_k-t_{k-1}$ and $k=1,2,...,M$. The procedure of calculating Hurst exponent contains 5 steps.

(1) Construct a mean variance series $X(j)$ based on $\{x(\tau_k)\}$,
\begin{equation}
	X(j)=\sum_{k=1}^{j}[x(\tau_k)-\langle x\rangle]
\end{equation}
where $j=1,2,...M$ and $\langle x(\tau_k)\rangle=\frac{1}{M}\sum_{k=1}^{M}x(\tau_k)$.

(2) Divide the series $X(t)$ into $M_s$ non overlapping subsequences, where $M_s=int[M/s]$. When $M$ is big enough, we can ignore the remainder.

(3) Calculate the detrended fluctuation function for every subsequences
\begin{equation}
	f_{\nu}(s)=\frac{1}{s}\sum_{k=1}^{s}\left\{X_{\nu}\left[(\nu-1)s+k\right]-X_{\nu}(k)\right\}^{2}
\end{equation}
Where $\nu=1, 2, ...N_s$. $X_{\nu}$ is the $\lambda$-order polynomial fitting function of subsequence $\nu$ by the least square method. Thus, we call this $DFA-\lambda$.

(4) The $DFA-\lambda$ fluctuation function is the mean square root of all subsequences detrended fluctuation function
\begin{equation}
	F(s)=\left(\frac{1}{N_s}\sum_{\nu=1}^{N_s}f_{\nu}(s)\right)^{1/2}
\end{equation}

(5) Calculating $F(s)$ for given different $s$, which satisfies
\begin{equation}
	F(s)=Cs^{H}
\end{equation}
where $C$ is a constant. Plotting $\log (F(s)/C)$ versus $\log(s)$ by the double logarithmic axes.
The slope of $\log (F(s)/C)$ versus $\log(s)$ is Hurst exponent.
The Hurst exponent can be written as
\begin{equation}\label{HEp}
	H=\frac{\log(F(s)/C)}{\log s}
\end{equation}

Therefore Hurst exponent can be an index describing the environment of the stochastic time series. When $H>0.5$, the environment shows a tendency, which implies that investors could earn the speculate premium by using the trend following strategy like the momentum trading strategy. When $H<0.5$, the environment oscillates, which implies that investors could use the trend reversal strategy like contrarian trading strategy to optimize their returns. The Hurst exponent plays a signal role to indicate the arbitrary premium. Many practical investigations demonstrate that the Hurst exponent of Chinese stock index sequences is approximately $0.6$,\cite{Shi,Han} which implies there exists chance to gain the arbitrary premium from the Chinese stock market.

	% =========================
	% 参考文献
	% =========================
%	\bibliographystyle{unsrt}
%	\bibliography{references}
\bibliography{apssamp}

\begin{thebibliography}{99}
\bibitem{De}
F. M. De Bondt, R. Thaler. Does the stock market overreat? TheJournal of Finance,
40(3),793-805(1985)¡£

\bibitem{Loughran}
T. Loughran, J. R. Ritter. The new issues puzzle. The Journal of Finance,  50(1),
23-51(1995).

\bibitem{Chen}
N.-F. Chen, R. Roll and S. A. Ross. Economic forces and the stock market. The Journal of
business,  59(3) 383-403(1986).

\bibitem{ANTEGNA}
ANTEGNA R N, STANLEY H E. Scaling behavior in the dynamics of an economic index[J]. Nature, 376 46-49(1995) .

\bibitem{Peters}
E. E. Peters. Fractal market analysis: applying chaos theory to investment and economics. A Wiley Finance Edition,  39-53(1994).

\bibitem{Jegadeesh}
N. Jegadeesh, S. Titman. Profitability of momentum strategies: an evaluation of alternati-
ve explanations. The Journal of Finance, 56(2)  699-720(2001).

\bibitem{Gebhardt}
W. R. Gebhardt, S. Hvidkjaer and B. Swaminathan. Stock and bond market interaction: Does
Momentum spill over? Journal ofFinancial Economics,  75(3) 651-690(2005).

\bibitem{Drew}
M. E. Drew. Do momentum strategies work? Australian evidence. Managerial Finance,
33(10),772-787 (2007).

\bibitem{Wei}
Wei Chen, Shinong Wu. Studies of long-term memory on Chinese stock markets. CONTEMPORARAY FINANCE and ECONOMICS,  24(6),  49-51(2003).

\bibitem{Gao}
Hongbin Gao, Jin Pan, Hongmin, Chen. Volatility's Hurst expoment of Chinese security markets[J]. Journal of Donghua University ,  27(4),  22-25(2001).

\bibitem{Markowitz}
Markowitz, H.M. March,  "Portfolio Selection". The Journal of Finance. 7 (1),  77-91(1952).

\bibitem{Peters}
Peters E E. Fractal Maket Analysis: Applying Chaos Theory to Investment and Economics[M].
Hoboken,HJ: Wiley-Blackwell, (1994).

\bibitem{Peters2}
PETERS E E. Chaos and order in the Capital Markets: A New View of Cycles, Prices, and
Market Volatility[M]. Hoboken,HJ: Wiley-Blackwell, (1996).

\bibitem{Shi}
Yongdong Shi, Yonggang Zhao, Chinese stock markets'fractal structural-Practical studies on stocks'long term memory.[J]. Mathematics in Practice and Theory, 09(2006).

\bibitem{Han}
Guowen Han, Haibo, Han. Fractal market theory and practical analysis on Chinese security market. Journal of Beijing Jiaotong University,  01(2008).

\bibitem{Clark}

Jack Clark Francis, Dongcheol Kim, Modern Portfolio Theory: Foundations, Analysis, and
New Developments, John Wiley and Sons Inc, (2013).
	
\end{thebibliography}
% Produces the bibliography via BibTeX.

\end{document}